PERSPECTIVE ARTICLE

# Advancing Drug Resistance Research Through Quantitative Modeling and Synthetic Biology

Kevin Farquhar[1], Harold Flohr[2], and Daniel A. Charlebois[2,*]
[1]Precision for Medicine, Houston, TX, 77054, USA
[2]Department of Physics, University of Alberta, Edmonton, AB, T6G-2E1, Canada
*Corresponding Author: dcharleb@ualberta.ca

## Abstract

Antimicrobial resistance is an emerging global health crisis that is undermining advances in modern medicine and, if unmitigated, threatens to kill 10 million people per year worldwide by 2050. Research over the last decade has demonstrated that the differences between genetically identical cells in the same environment can lead to drug resistance. Fluctuations in gene expression, modulated by gene regulatory networks, can lead to non-genetic heterogeneity that results in the fractional killing of microbial populations causing drug therapies to fail; this non-genetic drug resistance can enhance the probability of acquiring genetic drug resistance mutations. Mathematical models of gene networks can elucidate general principles underlying drug resistance, predict the evolution of resistance, and guide drug resistance experiments in the laboratory. Cells genetically engineered to carry synthetic gene networks regulating drug resistance genes allow for controlled, quantitative experiments on the role of non-genetic heterogeneity in the development of drug resistance. In this perspective article, we emphasize the contributions that mathematical, computational, and synthetic gene network models play in advancing our understanding of antimicrobial resistance to discover effective therapies against drug-resistant infections.

**Keywords**: Antimicrobial resistance, gene regulatory networks, mathematical modeling, non-genetic heterogeneity, stochastic gene expression, synthetic biology.

**Introduction**

Antimicrobial resistance (AMR) is an emerging health crisis that is undermining modern medicine (World Health Organization, 2014). AMR arises when bacteria, fungi, viruses or other microbes no longer respond to the antimicrobial drugs used to treat them. As of 2016, 700,000 deaths per year are attributed to AMR (O'Neill and The Review on Antimicrobial Resistance, 2016). If unmitigated, it is estimated that by 2050, AMR will kill 10 million people per year globally and result in a cumulative lost global production cost of 100 trillion USD. Though it has been argued that these figures may be over-estimates (de Kraker et al., 2016), there is undoubtedly a large and increasing clinical and public health burden associated with AMR. Drug resistance during chemotherapy also continues to be the major limiting factor for successfully treating patients with cancer (Vasan et al., 2019). In order to mitigate drug resistance, we need to establish new quantitative tools to study the drug resistance process, to discover new drugs, and to develop novel treatment strategies that extend the "lifespan" of antimicrobial and chemotherapy drugs.

It is well established that drug resistance can develop through genetic mutation (Figure 1A) that causes a permanent change in a micro-organism's DNA or through the acquisition of a drug resistance gene (e.g., horizontal gene transfer that occurs in bacteria) (Ochman et al., 2000). More recently, research has uncovered a new form of non-genetic stress resistance that can arise from fluctuations in gene expression in clonal cell populations (Figure 1B) (Fraser and Kærn, 2009; Geiler-Samerotte et al., 2013; van Boxtel et al., 2017); this, for example, includes the non-genetic drug resistance associated with the increased expression of genes that encode efflux proteins that pump antimicrobial drugs out of pathogenic yeasts such as *Candida glabrata* (Ben-Ami et al., 2016) and *Cryptococcus neoformans* (Mondon et al., 1999). Targeting this phenomenon will be important for mitigating AMR, as some non-genetically drug-resistant pathogens are not easily detected by standard laboratory tests (Sears and Schwartz, 2017) and non-genetic drug resistance may be associated with the failure of antimicrobial therapies (Ben-Ami et al., 2016) and chemotherapies (Brock et al., 2009). Non-genetic drug resistance can be modulated and enhanced by the structure of a gene regulatory network (Charlebois et al., 2014; Inde and Dixon, 2018; Camellato et al., 2019). The emerging paradigm is that drug resistance is a multi-stage process and that acute, non-genetic drug resistance can facilitate the evolution of permanent, genetic drug resistance (Figure 1C) (Brock et al., 2009; Charlebois et al., 2011; Bodi et al., 2017; Farquhar et al., 2019). For instance, it is known that mutations in PDR1, a gene that regulates PDR5 in the pleiotropic drug resistance (PDR) network in yeast (Figure 2A), can cause full resistance to the antifungal drug fluconazole (Ferrari et al., 2009). Though, more research is needed to elucidate the interplay between non-genetic and genetic forms of drug resistance.

Mathematical models of drug resistance have been used for over three decades (Lavi et al., 2012); many older mathematical studies were based on ABC (ATP-binding cassette) transporters, such as the PDR5 gene that is regulated by PDR1 in the PDR network, as the main mechanism of resistance. These models are now beginning to include more contemporary knowledge of AMR mechanisms and incorporate how drug resistance gene networks function and evolve during treatment (Charlebois et al., 2014; Farquhar et al., 2019). Mathematical models have the potential to predict the effectiveness of various treatment strategies, such as using combination drug therapies to overcome AMR (Baym et al., 2016), and can guide laboratory experiments by identifying experimental targets and by narrowing down the immense number of ways that antimicrobial drugs can be applied. Additionally, these models can elucidate mechanisms underlying the development of AMR (e.g., Farquhar et al., 2019) and predict AMR from experimental data (Arepyeva et al., 2017).

Synthetic biology is rapidly becoming part of the solution to many of our needs in medicine, agriculture, and energy production (El Karoui et al., 2019). A particularly promising application is to genetically engineer micro-organisms to carry synthetic gene networks to study AMR in a more quantitative, controlled, and efficient manner than has been possible using traditional ("natural" or non-genetically modified) model micro-organisms (González et al., 2015). At present, it is extremely challenging to develop and experimentally validate mathematical models using pathogens, where drug resistance genes have evolved to be highly connected to the host genome; for instance, the expression of MDR1/p-glycoprotein (responsible for multiple drug resistance (MDR) of tumours to chemotherapy; Gottesman et al., 2002) is regulated by a multitude of factors, making it difficult to quantitatively study how non-genetic mechanisms may contribute to AMR and drug resistance in cancer (Camellato et al., 2019). Furthermore, unlike synthetic drug resistance networks, native resistance networks are still not known completely. The design of synthetic gene networks is a model-guided process (Sakurai and Hori, 2017) and these networks are constructed to mimic natural drug resistance networks using techniques from genetic engineering (Cameron et al., 2014; Bartley et al., 2017).

**Mathematical Modeling of Non-Genetic Antimicrobial Resistance**

*Modeling Non-Genetic Gene Expression Heterogeneity in Drug Resistance*

Early work on non-genetic drug resistance focused on the amplitude of fluctuations or "noise" in the expression of drug resistance genes. Models predicted that low gene expression noise would be beneficial under low drug concentrations and that high gene expression noise would be beneficial under high drug concentrations (Figure 1B, inset) (Blake et al., 2006; Zhuravel et al., 2009); these predictions were confirmed experimentally in the same studies. Subsequently, a more general theoretical framework was developed that incorporated the frequency of gene expression noise, as well as the amplitude (Charlebois et al., 2011). Importantly, using this quantitative framework it was hypothesized that drug resistance can develop independently of mutation, provided that the fluctuation timescales are sufficiently long. Cell population models (Arino and Kimmel, 1993; Henson, 2003; Charlebois and Balázsi, 2019) have also been used to incorporate the multi-scale nature of AMR. For instance, a stochastic model of gene expression was combined with a population simulation algorithm to computationally investigate the evolution of gene expression noise (Charlebois, 2015).

*Modeling Drug Resistance Networks in Microbes and Mammalian Cells*

Mathematical models have been used to investigate the effect gene network structures or motifs have on AMR. For instance, it was shown computationally that gene network motifs can enhance drug resistance by modulating non-genetic gene expression variability within a cell population (Charlebois et al., 2014). Charlebois et al. showed that feedforward and positive feedback loops, the same network motifs that have been found to be imbedded in some gene networks regulating AMR in pathogenic yeast (Kolaczkowski et al., 1998) and human cancer cells (Misra et al., 2005), enhance drug resistance *in silico*. This new understanding of how gene network structure regulates AMR opens up new lines of research and identifies new potential therapeutic targets (e.g., targeting regulator genes in the network, rather than the drug resistance genes they control) against drug-resistant pathogens and cancers to be investigated experimentally.

Mathematical modeling and computer simulations have been used to predict how drug efflux pump proteins affect gene network function and fitness in prokaryotic and eukaryotic organisms. In Langevin et al. it was found experimentally that the cellular fitness benefit of AcrAB-TolC, a well-known multi-drug resistance pump in *E. coli*, depended on the rate of stress induction; fits to data allowed the fitness benefit that the pumps conferred under different stress induction rates to be accurately predicted by mathematical models (Langevin et al., 2018). In another study, it was predicted that incorporating negative feedback via drug efflux pumps into synthetic gene networks can increase the response of the gene network at low antibiotic inducer concentrations (Diao et al., 2016). This prediction was confirmed experimentally in the same study using synthetic gene networks in *S. cerevisiae* and was found to be the result of reduced regulator gene expression.

In Farquhar et al., the authors developed a stochastic population dynamics model to infer mechanisms for drug resistance in mammalian cells (Farquhar et al., 2019). The stochastic population model predicted that gene network motifs facilitate the development of acute drug resistance and that non- or slow-growing subpopulations of "persister-like" cells (see Brauner et al., 2016, Rosenberg et al., 2018, and Berman and Krysan, 2020 for the distinction between "tolerance", "heteroresistance" or "persistence", and "resistance") that do not succumb are critical reservoirs for the development of fast growing, heritably drug-resistant mutants enabling longer-term drug resistance. This study compliments previous work in bacteria that demonstrated that antibiotic tolerant non- or slow-growing mutant cells precede the developed genetic drug resistance during intermittent antibiotic exposure (Levin-Reisman et al., 2017). The persistence phenotype (e.g., Kussell et al., 2005) and stochastic phenotype switching (e.g., Acar et al., 2008) have also been investigated in mathematical models and experiments on genetically engineered micro-organisms and found to affect fitness in fluctuating environments.

Ultimately, mathematical and computation models of AMR must be validated by performing quantitative drug resistance experiments; genetically engineered cells that harbor synthetic gene networks controlling the expression of drug resistance genes is proving to be an effective experimental model system.

**Synthetic Drug Resistance Gene Networks & Antimicrobial Resistance Experiments**

Genetic engineering techniques are used to synthetize and combine DNA to build synthetic gene networks or "circuits" (Cameron et al., 2014) that control drug resistance genes. Common methods used to engineer synthetic gene networks include recombinant molecular cloning, Gibson assembly (Gibson et al., 2009; Santos-Moreno and Schaerli, 2019), and CRISPR-Cas9 gene editing (Nissim et al., 2014). Cell-to-cell heterogeneity may cause unexpected deviation from intended synthetic gene circuit behavior (Beach et al., 2017). However, statistical tools can enhance the design process and reliability of synthetic gene networks (Sakurai and Hori, 2017). With proper design, synthetic gene networks can be precisely tuned to control gene expression mean and noise levels using chemical inducers that do not adversely affect the micro-organisms harboring these networks.

*Synthetic Antimicrobial Resistance Gene Networks*

Synthetic gene networks have been engineered to regulate drug resistance and have been shown to serve as well-characterized models of natural stress response modules in evolution experiments (González et al., 2015; Bodi et al., 2017; Farquhar et al., 2019; Gouda et al., 2019). A synthetic two-gene positive feedback network controlling the Zeocin antibiotic resistance gene enables bi-stable gene expression was constructed in *S. cerevisiae* (Nevozhay et al., 2012). In this work, a computational approach based on stochastic cellular movement in gene expression space was used to predict cell population fitness of low- and high-expressing subpopulations. The authors found an optimum on the fitness landscape that balances the costs and benefits of expressing a drug resistance gene in various experimental antibiotic inducer and drug conditions. In a subsequent microbial evolution study using the same positive feedback yeast strain, it was found that the synthetic gene network was fine-tuned by evolution to modulate the network's noisy response and optimize fitness via specific "intra-circuit" and "extra-circuit" DNA mutations (González et al., 2015), which can lead to loss of gene circuit function that can be regained in certain conditions under drug selection (Gouda et al., 2019). The study by Gouda et al. also suggests that slow growth due to antibiotics may allow cells to access hidden drug-resistant states prior to genetic changes. Computational models based on fitness and gene expression properties have been developed to predict specific aspects of evolutionary dynamics (including the speed at which the ancestral genome disappears from the population and the types and number of mutant alleles that establish in each experimental condition) in different inducer and drug conditions (González et al., 2015). These computational models were validated in the same studies by microbial evolution experiments on the genetically engineered positive feedback yeast strain (Nevozhay et al., 2012; González et al., 2015; Gouda et al., 2019).

Genetically engineered networks have also been designed to control the expression of genes that encode efflux proteins that lead to AMR. Diao et al. used synthetic negative feedback gene networks, inducible by the antibiotic doxycycline, to regulate the expression of PDR5 (Diao et al., 2016). This study found that the addition of a second layer of negative feedback (resulting from pumping doxycycline out of the cell by the PDR5 protein) altered the dose-responses of the original gene circuits in a manner that was predictable by mathematical modeling. In another study, Camellato et al. engineered a synthetic gene network in yeast to mimic the PDR5 and MDR1 networks that underly multi-drug resistance in yeast and human breast cancer cells (Figure 2B) (Camellato et al., 2019). In agreement with computational predictions made years earlier (Charlebois et al., 2014), the authors found that feedforward and positive feedback network motifs enabled rapid, self-sustained activation of gene expression leading to enhanced cell survival in the presence of a cytotoxic drug. It has been proposed that activating the expression of genes that encode multi-drug resistance efflux pump proteins in the absence of antibiotic pressure may allow susceptible bacteria to outcompete resistant bacteria, which normally down-regulate the expression of resistance genes in conditions without antibiotics to eliminate the associated fitness cost (Wang et al., 2019).

*Synthetic Drug Resistance Gene Circuits in Mammalian Cells*

To experimentally investigate the role of non-genetic heterogeneity in cancer drug resistance, it is imperative to precisely control the non-genetic heterogeneity that can drive adaptation to chemotherapeutic agents. Synthetic gene circuits integrated in mammalian cells can be designed to precisely control gene expression noise for drug resistance genes, while keeping the mean expression levels the same (Figure 2F) (Aranda-Díaz et al., 2017; Farquhar et al., 2019). This approach allows synthetic gene circuits to separate key biological variables contributing to resistance from other confounding variables like mean expression and genetic background.

In Chinese Hamster Ovary (CHO) cells with a Flp-In landing pad (Wirth and Hauser, 2004), Farquhar et al. designed, constructed, and integrated into the landing pad a mammalian negative feedback (mNF) synthetic gene circuit (Figure 2C) (Farquhar et al., 2019) based on a humanized tetracycline repressor (hTetR) regulator gene (Nevozhay et al., 2013); the mNF circuit demonstrated doxycycline-inducible expression of a purmoycin drug resistance gene (PuroR) with low gene expression noise (Figure 2E). Highlighting the advantages of mathematical modeling in synthetic biology, the mNF circuit was originally developed from a derivative gene circuit transferred from yeast in a previous study that applied modeling to modify the design of the circuit through multiple iterations, leading to increased fold change and minimal gene expression noise (Nevozhay et al., 2013). Complementing the low noise drug resistance synthetic gene circuit, Farquhar et al. also constructed a mammalian positive feedback (mPF) gene circuit (Figure 2D), integrated into the same CHO genomic locus as the mNF circuit, leading to doxycycline-inducible expression of PuroR with high levels of gene expression noise (Figure 2G) (Farquhar et al., 2019). When inducing the two circuits in mammalian CHO cells to express the same PuroR mean expression level (Figure 2F) and treating the cells with various concentrations of puromycin, the authors found that adaptation to low concentrations of drug was favored for the mNF circuit with low gene expression noise. On the other hand, high gene expression noise from the mPF circuit facilitated puromycin resistance under high drug concentrations. This validated the approach to investigating drug resistance and noise in mammalian cells using synthetic gene networks, which were required to decouple noise from mean drug resistance gene expression in isogenic cells; this approach could also help to further elucidate the role of rare-cell expression and drug-induced reprogramming in mammalian drug resistance (Shaffer et al., 2017).

DNA sequencing of the gene circuits after adaptation and monitoring expression and survival after temporary removal of drug helped reveal adaptation mechanisms (Farquhar et al., 2019). In the mNF circuit, the self-repression from the hTetR regulator tended to break down through intra-circuit mutations, leading to higher PuroR expression and irreversible resistance to puromycin. In the mPF circuit, no intra-circuit mutations were found despite PuroR expression levels remaining elevated above pre-treatment mean expression levels, which was reversible and led to re-sensitization to puromycin after inducer removal. By using synthetic gene networks containing a drug resistance gene in isogenic mammalian cells, Farquhar et al. addressed a long-standing question regarding how non-genetic heterogeneity could lead to initial cell survival during chemotherapy which then facilitates the development of genetic drug resistance in cancer (Brock et al., 2009).

**Discussion**

A new interdisciplinary field of research is emerging that combines multi-scale quantitative models with synthetic biology to rationally design gene networks using engineering principles for AMR research. One important goal is to use these models to predict the effects of non-genetic drug resistance on the evolution of genetic drug resistance. Another important goal is to advance pharmaceutical and clinical AMR research by investigating new "resistance proof" antimicrobial compounds and novel therapeutic treatment strategies.

Moving forward, a challenge that must be addressed is how to adapt the mathematical models and translate the experimental discoveries made using synthetic systems to pathogens with complex and highly interconnected gene regulatory networks. More research on pathogenic micro-organisms and mammalian cells is needed to elucidate the underpinnings of non-genetic resistance at the molecular and single-cell levels. This research will be critical to fully understand how non-genetic and genetic mechanisms interact in the development of drug resistance, and to discover effective strategies that target acute non-genetic drug resistance to alleviate the development of permanent genetic drug resistance in infectious diseases and cancers. Several promising approaches include synergistically using noise-enhancing compounds to reactivate latent HIV to increase sensitivity to existing antiviral drugs (Dar et al., 2014), using combined drug treatment regimens to target non-proliferating *M. tuberculosis* persisters to reduce treatment times (Zhang et al., 2012), eliminating bacterial persisters using engineering approaches that target bacterial metabolism (Allison et al., 2011), and the use of epigenetic modifiers in combination with targeted therapies to reduce the ability of a cancerous cell to switch phenotypes to acquire a drug-resistant state (Salgia and Kulkarni, 2018).

Overall, combining mathematical models and synthetic gene networks is leading to new quantitative model systems for drug resistance research, which are desperately needed to advance our fundamental understanding of the multi-stage drug resistance process. Ultimately it remains to be seen how discoveries made using these quantitative model systems will translate to pathogens and cancer. However, the potential of this new area of research to help mitigate the socio-economic costs of drug resistance warrants its relentless pursuit.

**Author Contributions**

DC conceptualized and supervised the study. KF, HF, and DC all contributed to the literature review. HF and DC developed the figures. KF and DC wrote the manuscript.

**Funding**

Precision provided indirect funding support for this paper by way of employment of KF. No additional funding was provided for this paper by Precision or any other entity.

DC is supported in part by funding from the Social Sciences and Humanities Research Council (SSHRC) of Canada's New Frontiers in Research Fund (NFRFE-2019-01208).

**Conflict of Interest**

The authors declare that the research was conducted in the absence of any commercial or financial relationships that could be construed as a potential conflict of interest.

KF was employed by Precision for Medicine.

**Acknowledgments**

We are grateful to Prof. Gábor Balázsi for helpful comments on the manuscript. We thank Mr. Mark Igmen for assisting with the literature review.

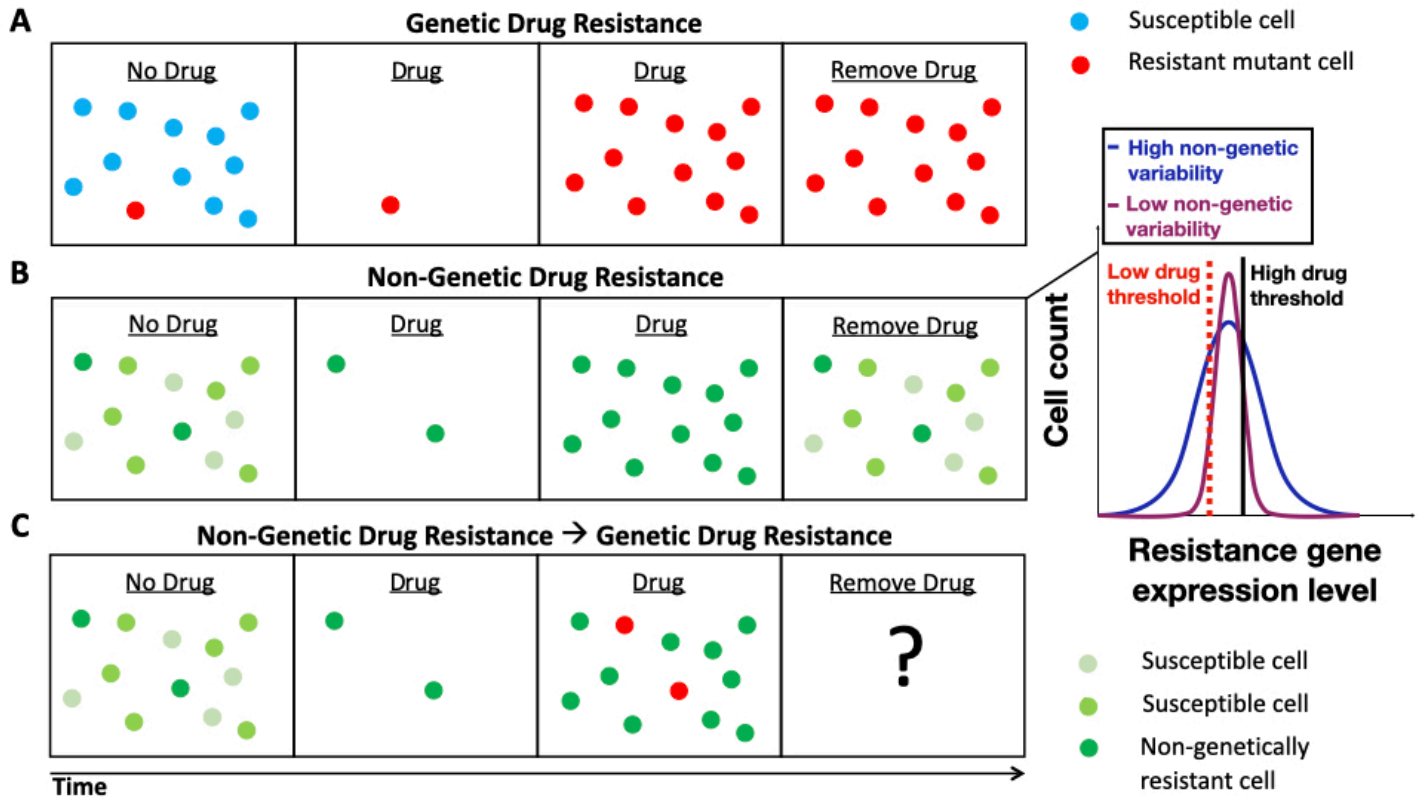

**Figure 1**. Schematic depicting the development of non-genetic and genetic drug resistance. **(A)** The development of genetic drug resistance via evolution by natural selection of a drug resistance mutation. **(B)** The development of non-genetic drug resistance in a clonal cell population via the selection cells with sufficiently long gene expression fluctuation timescales. The shade of green denotes the level of gene expression of a drug resistance gene inside the cell; lighter and darker shades of green represent lower levels and higher levels of gene expression, respectively. Inset illustrates gene expression histograms typically obtained from clonal cell populations with low and high levels of non-genetic, cell-to-cell variability. A high level of non-genetic variability is advantageous at high drug concentrations and a low level of non-genetic variability is advantageous at low drug concentrations (cells with resistance gene expression levels below a given drug threshold perish). **(C)** The evolution of longer-term, genetic drug resistance is facilitated by shorter-term, non-genetic drug resistance; the ultimate outcome will be determined based on the condition-dependent relative fitness of each subpopulation.

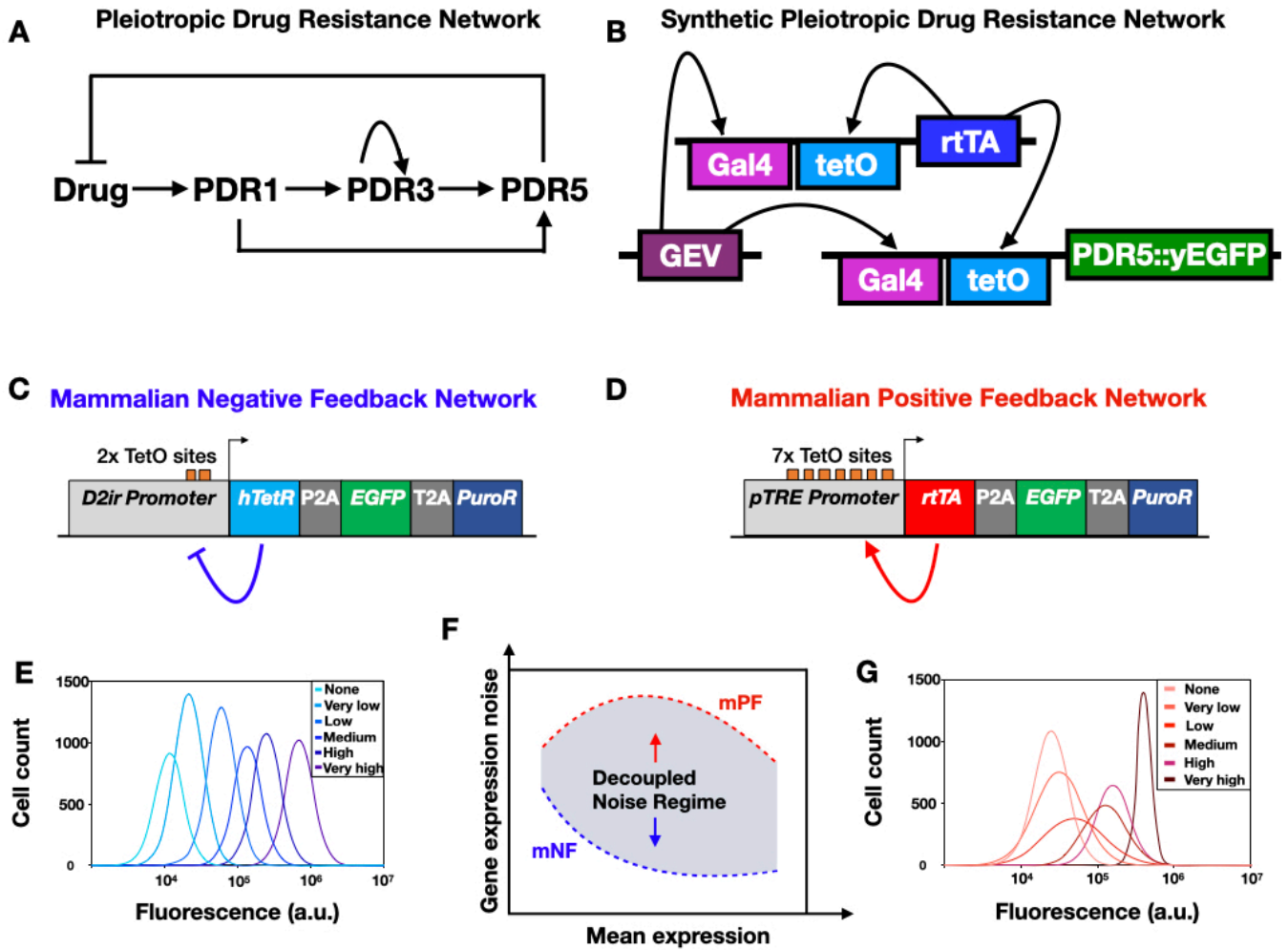

**Figure 2**. Synthetic gene networks engineered to mimic natural drug resistance gene networks and study the effects of non-genetic heterogeneity on AMR. **(A)** Yeast pleiotropic drug resistance (PDR) gene network. Imbedded in this gene network structure are positive feedback regulation (self-activation of PDR3), negative feedback regulation (efflux of gene network-inducing drug by PDR5), and feedforward regulation (PDR1 indirectly activates PDR5 through PDR3) and direct activation (PDR1 activates PDR5). **(B)** A synthetic gene network engineered to have the same network motif as the PDR network shown in **(A)**. Note that fluorescence reporter genes, such as yEGFP, are fused to drug resistance genes, such as PDR5, to enable experimental measurement. **(C)** Schematic of the mammalian negative feedback (mNF) gene network, which expresses the humanized tetracycline repressor (hTetR) gene, the puromycin resistance gene (PuroR), and the fluorescence reporter EGFP separated by self-cleaving 2A elements. **(D)** Schematic of the mammalian positive feedback (mPF) gene network, which expresses the reverse tetracycline transactivator (rtTA), PuroR, and EGFP separated by self-cleaving 2A elements. **(E)** Schematic of representative gene expression histograms obtained from a dose response of the mNF strain. **(F)** Decoupled noise regime which occurs when the mean gene expression noise amplitude for the mPF (high noise) and the mNF (low noise) gene networks is decoupled from mean gene expression. **(G)** Schematic of representative gene expression histograms obtained from a dose response of the mPF strain. Legends in **(E)** and **(G)** indicate the inducer (doxycycline) level for each distribution. Panel **(A)** was reproduced from Charlebois et al., 2014 with permission (https://doi.org/10.1103/PhysRevE.89.052708), panel **(B)** was adapted with permission from Camellato, 2018, and panels **(C)** to **(G)** were adapted from Farquhar et al., 2019, CC BY 4.0 (https://creativecommons.org/licenses/by/4.0/).